\documentclass[paper,notoc]{JHEP3}

\JHEPspecialurl{http://jhep.sissa.it/JOURNAL/JHEP3.tar.gz}

\usepackage{amsmath}
\allowdisplaybreaks[1]

\usepackage{epsfig,multicol}

\newcommand{\pt}{\mbox{$p_T$}}

\newcommand{\ptmiss}{\mbox{$p_T^{miss}$}}

\newcommand{\Gcs}{${\rm{GeV/}c^2}$}

\newcommand{\Gc}{${\rm{GeV/}c}$}
\newcommand{\Gcsit}{\ensuremath{\it{GeV/c}^2}}

\newcommand{\delphi}{\mbox{$\Delta \phi$}}
\newcommand{\ttbar}{\mbox{$t \overline{t} $}}

\newcommand{\fbs}{\mbox{$\rm{fb}^{-1}$}}

\newcommand\fverb{\setbox\pippobox=\hbox\bgroup\verb}
\newcommand\fverbdo{\egroup\medskip\noindent%
\fbox{\unhbox\pippobox}\ }
\newcommand\fverbit{\egroup\item[\fbox{\unhbox\pippobox}]}
\newbox\pippobox

\newcommand{\qqWW}{\mbox{$q \bar{q} \to WW$}}
\newcommand{\ggWW}{\mbox{$gg \to WW$}}
\newcommand{\etall}{\mbox{$\eta_{\ell \ell}$}}
\newcommand{\delphill}{\mbox{$\Delta \phi_{\ell \ell}$}}

\title{The process gg $\rightarrow$ WW as a background to the Higgs signal at the LHC}

\author{M.~D\"uhrssen, K. Jakobs and J.~J.~van der Bij\\
Institut f\"ur Physik, Albert-Ludwigs Universit\"at Freiburg, \\
         Hermann-Herder-Str. 3, 79104, Freiburg, Germany\\
E-mail: \email{Michael.Duehrssen@physik.uni-freiburg.de},
        \email{Karl.Jakobs@physik.uni-freiburg.de},
        \email{jochum@physik.uni-freiburg.de}}

\author{P.~Marquard\\
Institute of Particle Physics Phenomenology, Department of Physics, University of Durham,
        Durham, DH1 3LE, UK\\
E-mail: \email{Peter.Marquard@durham.ac.uk}}

\preprint{Freiburg-THEP-05/01\\DCPT/05/08\\IPPP/05/04\\hep-ph/0504006}

\abstract{The production of $W$ pairs from the one-loop gluon fusion process is studied.
Formulas are presented for the helicity amplitudes keeping the top mass finite, but
all other quark masses zero. The correlations among the leptons coming from the $W$ bosons 
are kept. The contribution of this background to the Higgs boson search in the 
$WW$ decay mode at the LHC is estimated by applying the cuts foreseen
in experimental searches using the PYTHIA Monte Carlo program. Kinematic distributions for the 
final state leptons are compared to those of the Higgs boson signal and of the 
\qqWW\ background. After applying final cuts, the $gg$  background is found to be large, 
at the level of 35\% of the $q \bar{q}$ background. 
The characteristics of the $gg$  background are very similar to those of the signal.
Therefore, an experimental normalization of this background component appears to be very 
difficult and the uncertainty must 
largely be determined by theory. As a result, the significance of a Higgs
signal in the $gg \to H \to WW$ mode at the LHC is reduced.
}

\keywords{One-loop, QCD, Higgs, Hadron Colliders}

\begin{document}

\section{Introduction}
The Standard Model is well established and in agreement with
all collider data. The only part of the model, not explored
so far, is the Higgs sector. Because this sector plays a distinguished role
in the theory, being responsible for the masses and mixings of all particles,
the search for the Higgs boson is one of the highest priorities
at the LHC. Within the Standard Model all properties of the Higgs
boson are fixed when its mass is known. From indirect limits the Higgs
boson is expected to have a mass in the range 114.4~\Gcs\ $< m_H <$ 246~\Gcs\ (95\% C.L.)
\cite{particlereview}.
When the Higgs mass is above the $Z$ pair threshold, it decays with a large branching
fraction into $Z$ bosons, that can be discovered in the ``golden"
$\ell^+\ell^+\ell^-\ell^-$ decay mode. As long as $m_H \gtrsim$ 130~\Gcs\ the decay into four
leptons can still be used. Within the region
155~\Gcs\ $< m_H <$ 170~\Gcs\ the $ZZ^*$ branching fraction goes through
a minimum, while the $WW^*$ decay mode opens up. 
In this mass range the $gg \to H \to WW \to \ell \nu \ \ell \nu$ \cite{dreiner} 
and the recently established vector boson fusion mode 
$qq \to qq H \to qq WW \to qq \ \ell \nu \ \ell \nu$ \cite{zeppenfeld-ww,atlas-vbf,cms-higgs}
have the largest discovery potential. 
Since two neutrinos appear in the final state, no 
invariant mass peak can be reconstructed. This implies that there is a large
background from the $W$ pair production process $q \bar q \rightarrow WW$.
This background can only be reduced by considering the specific differences
in distributions between the signal and background processes. The first important
difference is in the rapidity distributions of the final state leptons. 
The \qqWW\ background tends to be produced at a larger
rapidity, because of the harder distribution of the valence quarks, whereas
the signal comes from gluon fusion. The second difference lies in the lepton
correlations, since the $W$ bosons from the Higgs boson decay are in a scalar state.
It was shown in Ref.~\cite{dreiner} that it is possible, using these differences,
to strongly improve the signal-to-background ratio, in comparison to older
calculations that ignored the spin correlations \cite{will,barger}.
 To get a high confidence level for the signal, a normalization of the background is needed 
in a phase space region where the signal contribution is small.
Using Monte Carlo calculations an extrapolation into the signal region needs to be performed.
Hereby it is assumed that the shape of the distributions can be well predicted, even
when the normalization is uncertain.

The features that distinguish the $q \bar q$ background from the signal
are absent in the $gg \rightarrow WW$ process. Since the $gg$ process
can be quite large in the signal region, a calculation of this process is necessary.
The gluon fusion process has been studied in the literature before,
however, without accounting for correlations among the final decay products \cite{glover,kao}.
For the $ gg \rightarrow ZZ$ process, full calculations with correlations do exist
\cite{matsuura,zecher}. These programs can
easily be adapted to the production of $W$ bosons, as long as the quarks inside
the loop are degenerate, so as to describe the effect of the light quarks only.
For the top-bottom generation this is not correct and one has to separately calculate
this contribution to the helicity amplitudes. The formulas will be rather complicated
compared to the formulas of Ref.~\cite{bijglover} in the massless case.
Since a priori the presence of a heavy particle in the loop can affect the
distributions, one has to check this contribution. Ultimately we found
out, however, that the effect is negligible (10\%) as compared to the calculation
with only the light quarks.

The programs for the cross section calculation were interfaced with the
PYTHIA Monte Carlo program  \cite{pythia}, so
as to allow for a more realistic analysis. The comparison between the two backgrounds
was performed at leading order, since no K-factor for the gluon process is known.
Next-to-leading order (NLO) programs exist for the $q \bar q$ process
\cite{ohnemus1, ohnemus2,frixione,dixon1,dixon2,ellis}, however,
only rough estimates on the significance of the K-factors for the backgrounds 
can be made at present.

The outline of the paper is as follows: in Section 2
some theoretical issues are discussed, in Section 3 the simulation and the event 
selection is presented, in Section 4 results
for the different distributions are given and conclusions are drawn in Section 5.
The new formulas for the helicity amplitudes are given in the appendix.

\section{Theoretical issues}
\label{sec:theo}
For the process $gg\to W^+W^-$, which appears first at the one loop level,
a new calculation has been performed. In this calculation we included
the full dependence on the masses of the quarks inside the loop but
restricted ourselves to on-shell $W$ bosons and used the narrow-width
approximation instead. Only the box graphs contribute.
Off-shell $W$ bosons would only be interesting in the
threshold region, but otherwise lead to negligible corrections,
since there is no interference with the signal. The contribution of
the squared top-quark loop is of the order of 1\% while the
interference with light quarks gives a contribution of the order of 10\%.

To obtain the full
spin correlation we calculated the helicity amplitudes for $gg \to
W W$ and used spin matrices to combine them with the polarized
decay amplitudes for $W\to \ell \nu$. The calculation has been performed
using standard techniques \cite{vanOldenborgh:1989wn,
  Passarino:1978jh} to reduce the tensor integrals appearing in the
calculation and has been checked in several limits against known
calculations \cite{glover, matsuura, bijglover}. The remaining scalar integrals
have been evaluated numerically using the programs of Refs.~\cite{Hahn:1998yk,
  vanOldenborgh:1990yc}. The helicity amplitudes can be obtained from
the authors\footnote{email: Michael.Duehrssen@physik.uni-freiburg.de,
  Peter.Marquard@durham.ac.uk} together with the full Monte Carlo program.

Since no NLO calculation is available, the effect of the NLO
corrections can at present only be estimated.
Since the calculation is at leading order, the renormalization scale
dependence emerges only from the parton distribution functions (pdf) 
and the running of $\alpha_s$ and we chose $Q^2=\hat s /4$.

\section{Simulation and Event Selection}

In several studies \cite{dreiner,atlas-tdr,cms-ww}
it has been shown that the \qqWW, top pair and single top production constitute the principal
backgrounds in the Higgs boson search  in the $gg \to H \to WW \to \ell \nu \ \ell \nu$ 
mode. In the present study the \ggWW\ contribution of the background is compared to 
the \qqWW\ and \ttbar\ backgrounds. 
Both the Higgs boson signal and the 
background processes were generated with
PYTHIA 6.226~\cite{pythia} using the CTEQ5L \cite{cteq5l} structure function parametrization.
The decays of the $W$ bosons into all lepton flavours ($e \nu, \mu \nu$ and $\tau \nu$) have been
included. Since in the event selection only electrons and muons are considered, tau leptons 
only contribute if they decay leptonically. 
The $gg \to WW$ background was also generated with CTEQ5L and interfaced to PYTHIA
using the {\em Les Houches Accord} matrix element interface \cite{les-houches}. 

The signal selection is divided into two parts: a preselection is applied to
identify the final state and to reduce all backgrounds that contain no true $W$ pair
and in a final selection the signal is separated from the
dominant backgrounds as far as possible. It should be stressed that no simulation of 
detector effects was performed and cuts were applied based on particle level information
available in PYTHIA.

\subsection{Preselection}
\begin{enumerate}

\item The final state is triggered and selected by requiring exactly two isolated
leptons of opposite charge ($e$ or $\mu$) with $p_T(\ell_1)>$ 20~\Gc\ and $p_T(\ell_2)>10$~\Gc\
in the pseudorapidity interval $|\eta|<$ 2.5.
A lepton reconstruction efficiency of 90\% independent
of $p_T$ and $\eta$ was assumed. A lepton is considered to be isolated if
not more than 10 GeV of energy is carried by all particles within
$\Delta R = \sqrt{(\Delta \eta)^2 + (\Delta \phi)^2} <$~0.2 of the lepton.

\item The missing transverse momentum $p_T^{\rm miss}$ is defined as the sum of all neutrinos,
ignoring experimental smearing and resolution effects, and is required to be larger than
40~\Gc. This cut reduces all potential backgrounds with no
intrinsiv $p_T^{\rm miss}$ from neutrinos, {\em e.g.}, $q\bar{q} \to Z \to ee/\mu\mu$.

\item The modulus of the vector sum of the two leptons and the \ptmiss\ vector in the
transverse plane is required to be smaller than 60~\Gc. This requirement mainly reduces
the huge $t\bar{t}$ background.

\item Jets are reconstructed using a cone algorithm with $\Delta R=0.7$ and a particle seed of
$p_T>2$~\Gc. Furthermore, a jet is called a b-jet if a $b$ quark is found within
$\Delta R<0.4$ of the jet axis. For b-jets a detector reconstruction efficiency
of 60\% and a light jet mistag probability of 1\% is assumed.
To reduce the $t\bar{t}$ background, no jets with \pt\ above 40~\Gc\ are allowed within
$|\eta|<$ 4.5 and no b-jets with \pt\ above 20~\Gc\ within $|\eta|<$ 2.5.

\item Events containing $Z$ bosons are rejected by requiring that two same flavour
leptons of opposite charge may not have an invariant mass within 10~\Gcs\
of the $Z$ mass.

\item $Z \to \tau\tau$ events with $\tau \to \ell \nu \nu$ have 
the same final state as the signal, but the leptons
tend to be back-to-back. All events where the azimuthal opening angle \delphill\ between
the two leptons is larger than 3.0 are rejected. Using the collinear
approximation \cite{ellis-jochum}, it is possible to reconstruct the $\tau$ momenta.
If the collinear approximation is successful, \delphill\ is larger
than 2.0 and one of the $\tau$-decay leptons carries less than 50\% of the $\tau$ momentum,
the event is rejected.

\end{enumerate}

\subsection{Final Selection}
\begin{enumerate}

\item[7.] Some tails of the backgrounds are removed by requiring
$\Delta\eta(\ell_1,\ell_2) < 1.5$ and
$\etall :$~$=\eta(\vec{p}_{\ell_1}+\vec{p}_{\ell_2}) < 3.0$.

\item[8.] A cut on the pseudorapidity
of the sum of the two leptons, $\etall <$ 1.47,  
strongly suppresses the $q\bar{q} \to WW$ background.

\item[9.] The transverse mass $M_T$, defined as $M_T : = \sqrt{2\cdot p_T(\ell_1+\ell_2) \ 
p_T^{\rm miss} \cdot (1-\cos\phi_T)}$, where $\phi_T$ is the 
transverse opening angle between $\vec{p}_T^{\ \rm miss}$ and the sum of the two 
leptons $\vec{p}_T(\ell_1+\ell_2)$, 
is required to be in the range  $m_H$--30~\Gcs $< M_T < m_H$. Given the shapes of the 
transverse mass distributions for signal and backgrounds, this cut helps to 
increase the signal-to-background ratio for Higgs boson masses above 
150~\Gcs.

\item[10.] A final cut on $\delphill < 1.0$ exploits the main
difference between signal and background.

\end{enumerate}

\section{Results}

The acceptance for a Higgs boson signal with a mass of 170~\Gcs\ and for the various backgrounds
after the application of the successive cuts is summarized in Table \ref{t:acc}.
At the level of the basic acceptance cuts, {\em i.e.}, requiring two leptons and
the cut on the missing transverse momentum (cuts 1 and 2), the total background
is dominated by \ttbar\ production. This background can be largely suppressed by
applying jet vetos and a cut on the total transverse momentum of the
dilepton-\ptmiss\ system. After the preselection cuts, the $WW$ background is dominant
and sums up to a cross section of 603 fb. This is roughly five times larger than
the signal cross section. At that stage, the newly considered \ggWW\ contribution
amounts to only about 11\% of the total $WW$ background.

\begin{figure*}
\begin{minipage}{1.0\linewidth}
\begin{center}
\mbox{\epsfig{file=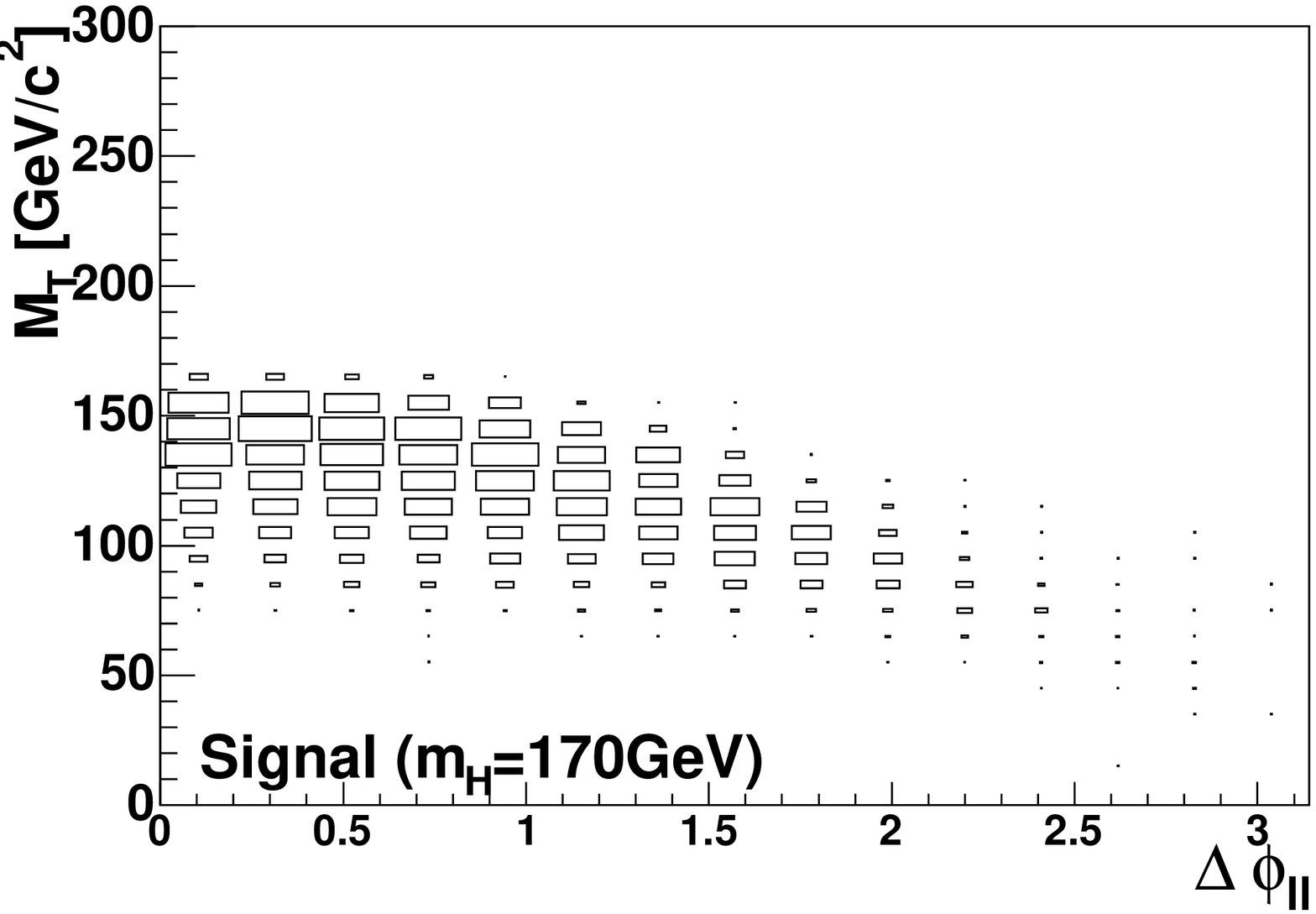,width=0.42\textwidth}}
\mbox{\epsfig{file=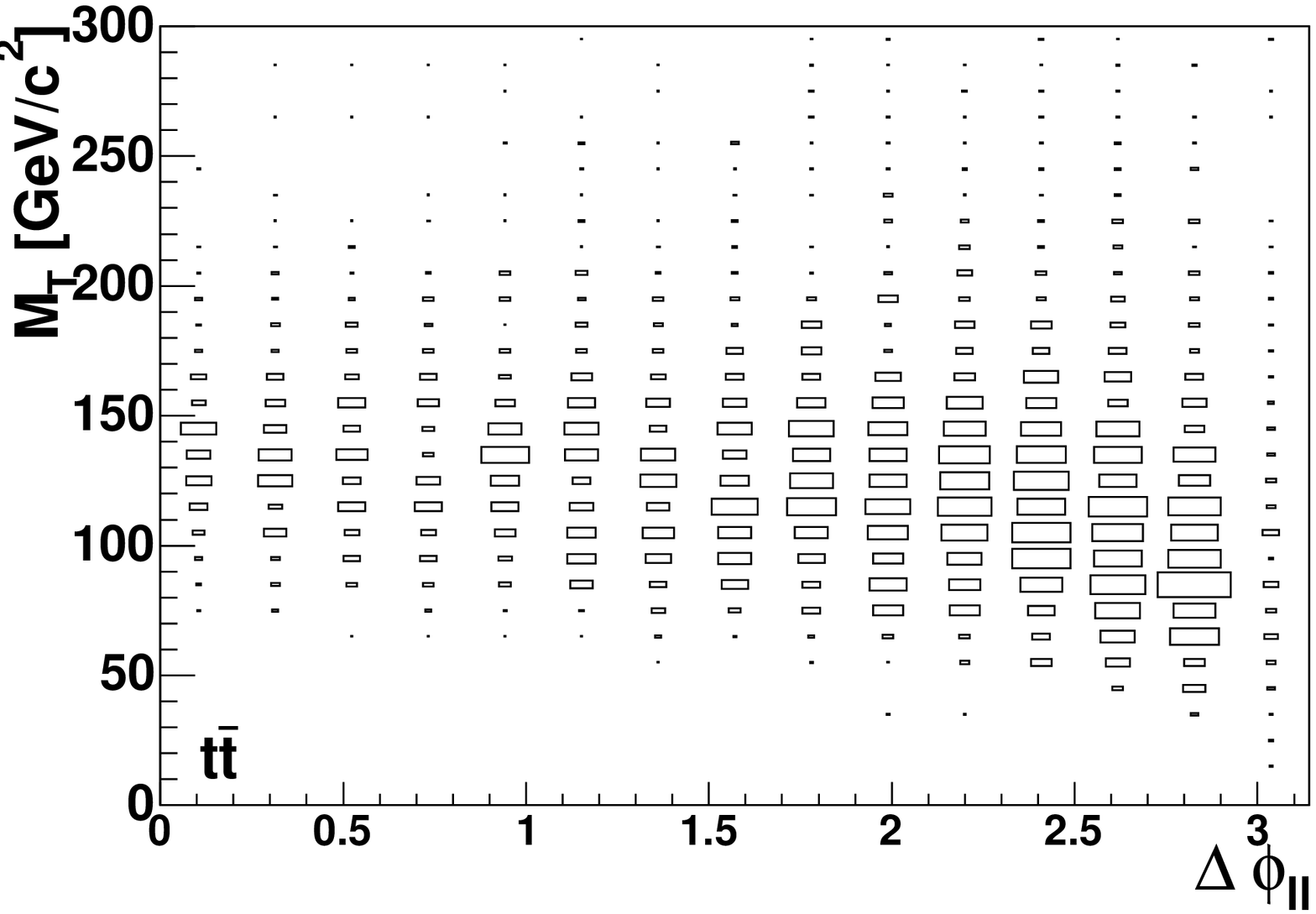,width=0.42\textwidth}}
\mbox{\epsfig{file=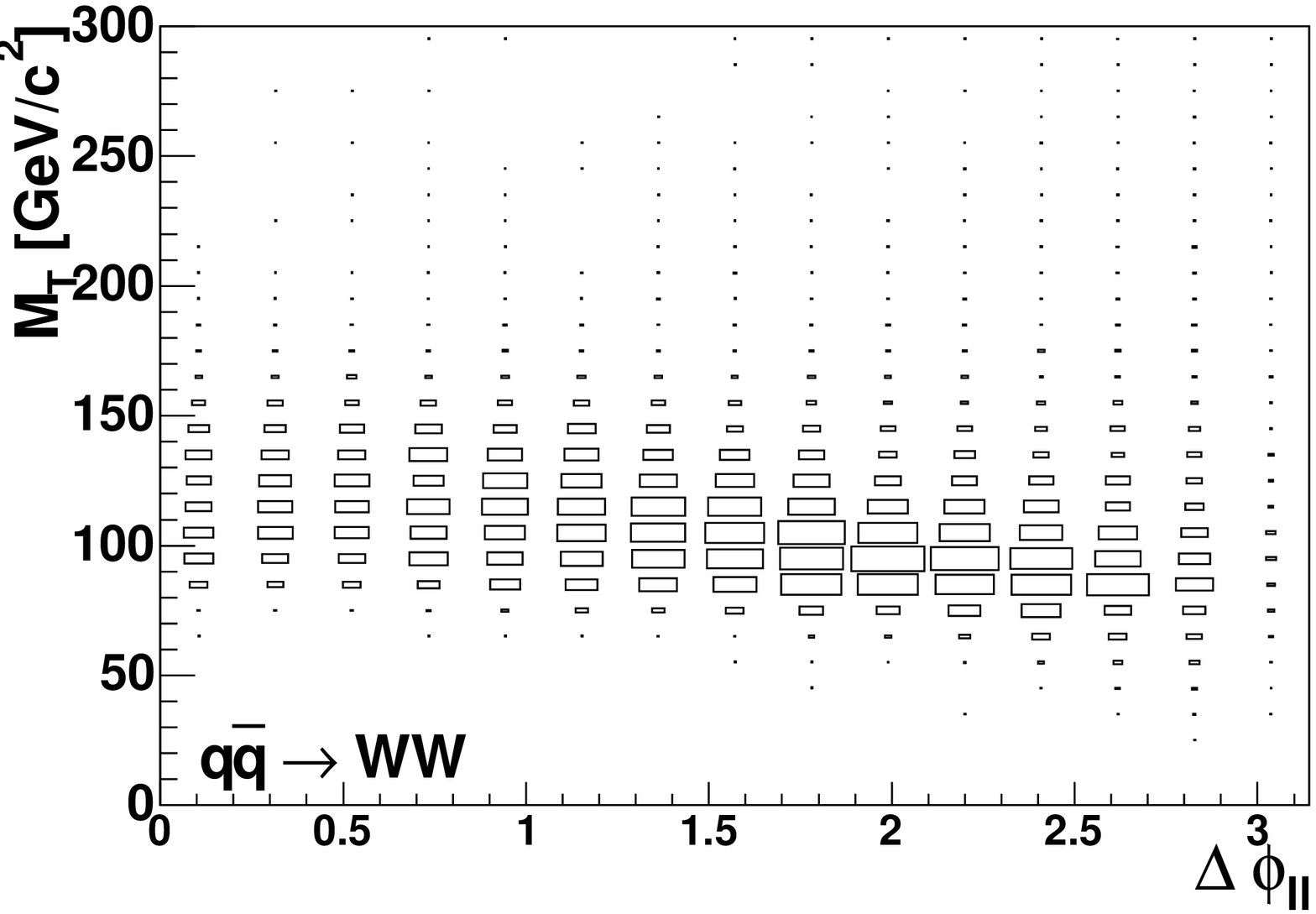,width=0.42\textwidth}}
\mbox{\epsfig{file=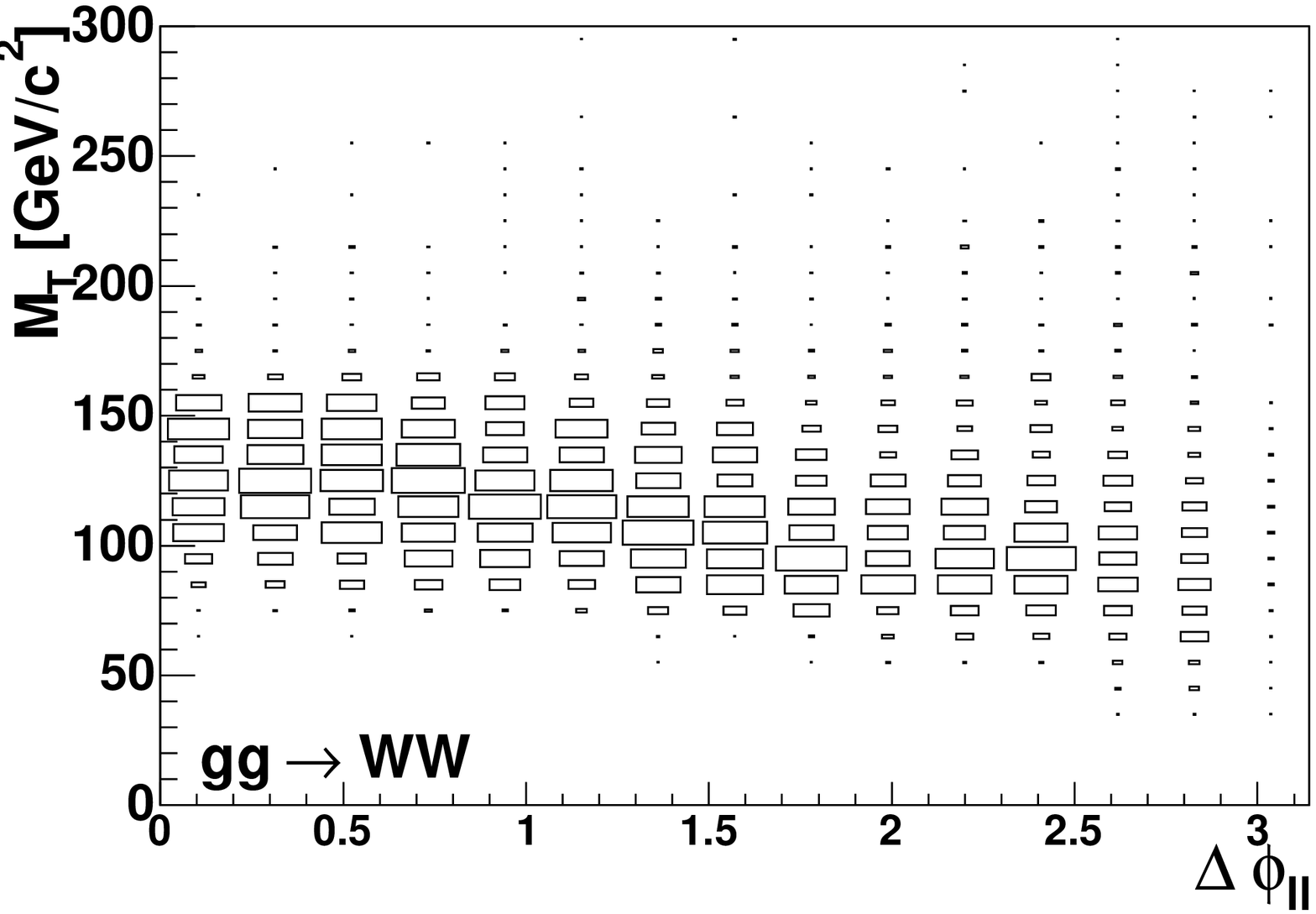,width=0.42\textwidth}}
\caption{\small \it
Distributions of the transverse mass $M_T$ versus \delphill\ for signal events
with $m_H$ = 170~\Gcsit\ and for the \ttbar, \qqWW\ and \ggWW\ backgrounds.}
\label{f:mt_delphi}
\end{center}
\end{minipage}
\end{figure*}

\begin{figure*}
\begin{minipage}{1.0\linewidth}
\begin{center}
\mbox{\epsfig{file=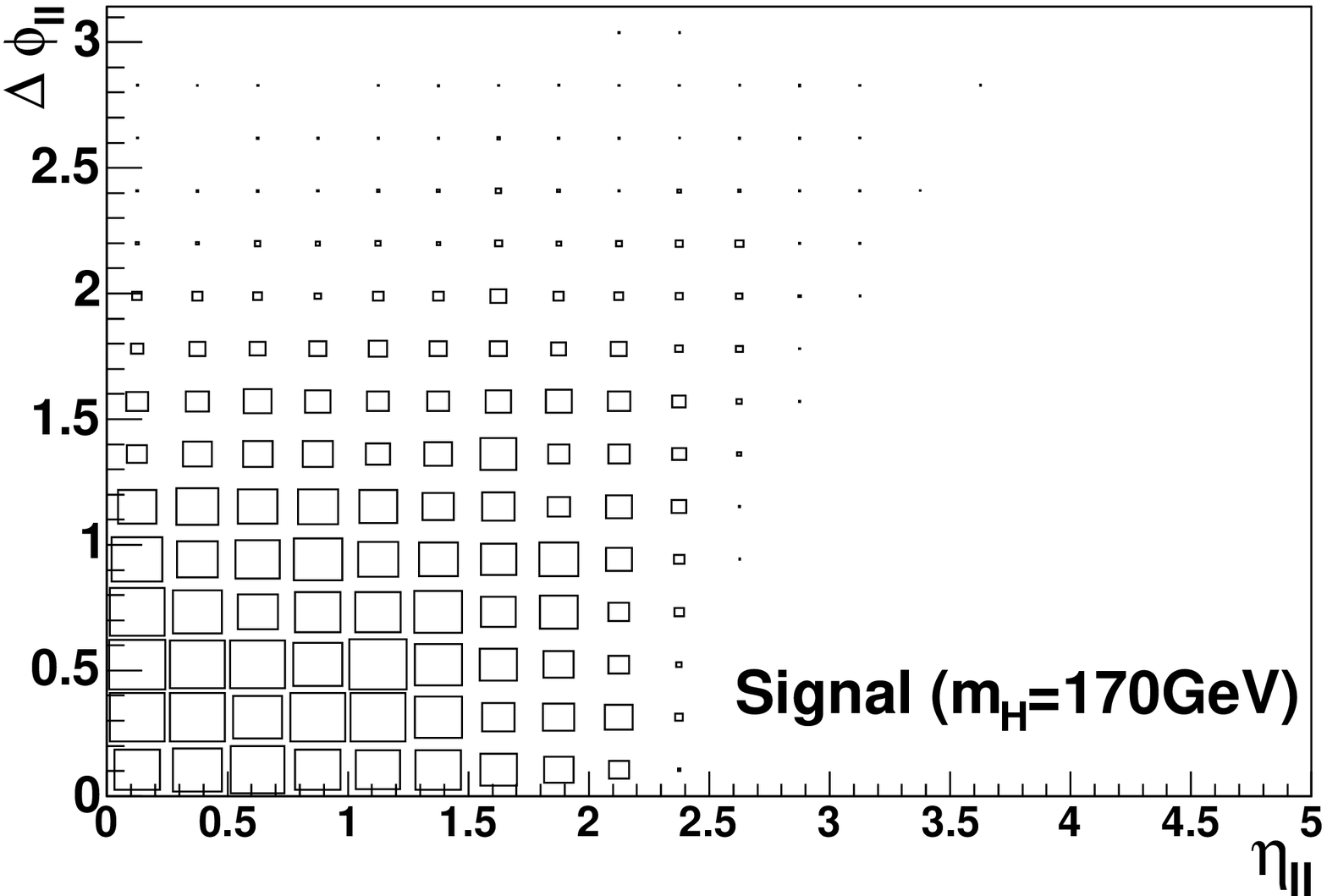,width=0.42\textwidth}}
\mbox{\epsfig{file=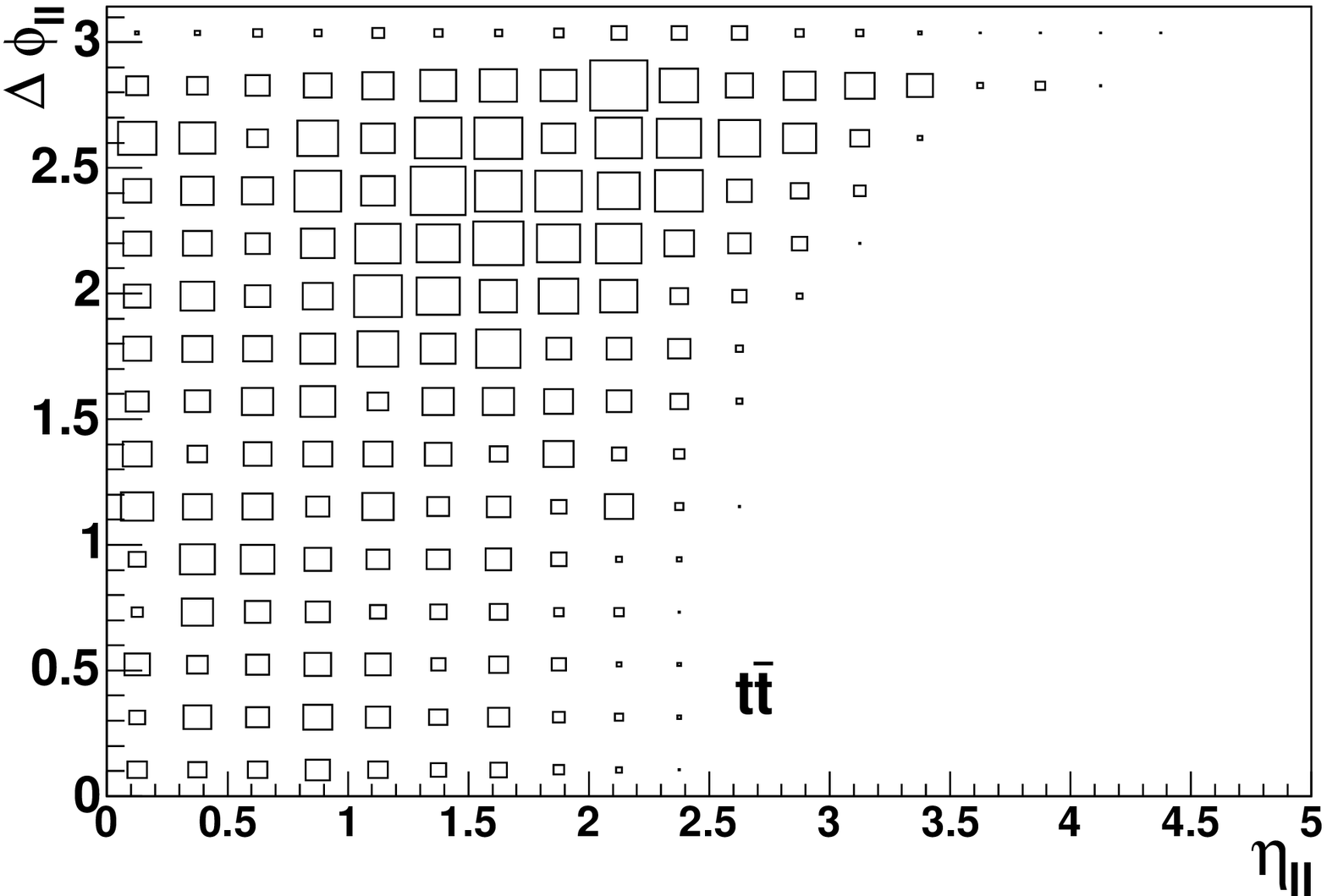,width=0.42\textwidth}}
\mbox{\epsfig{file=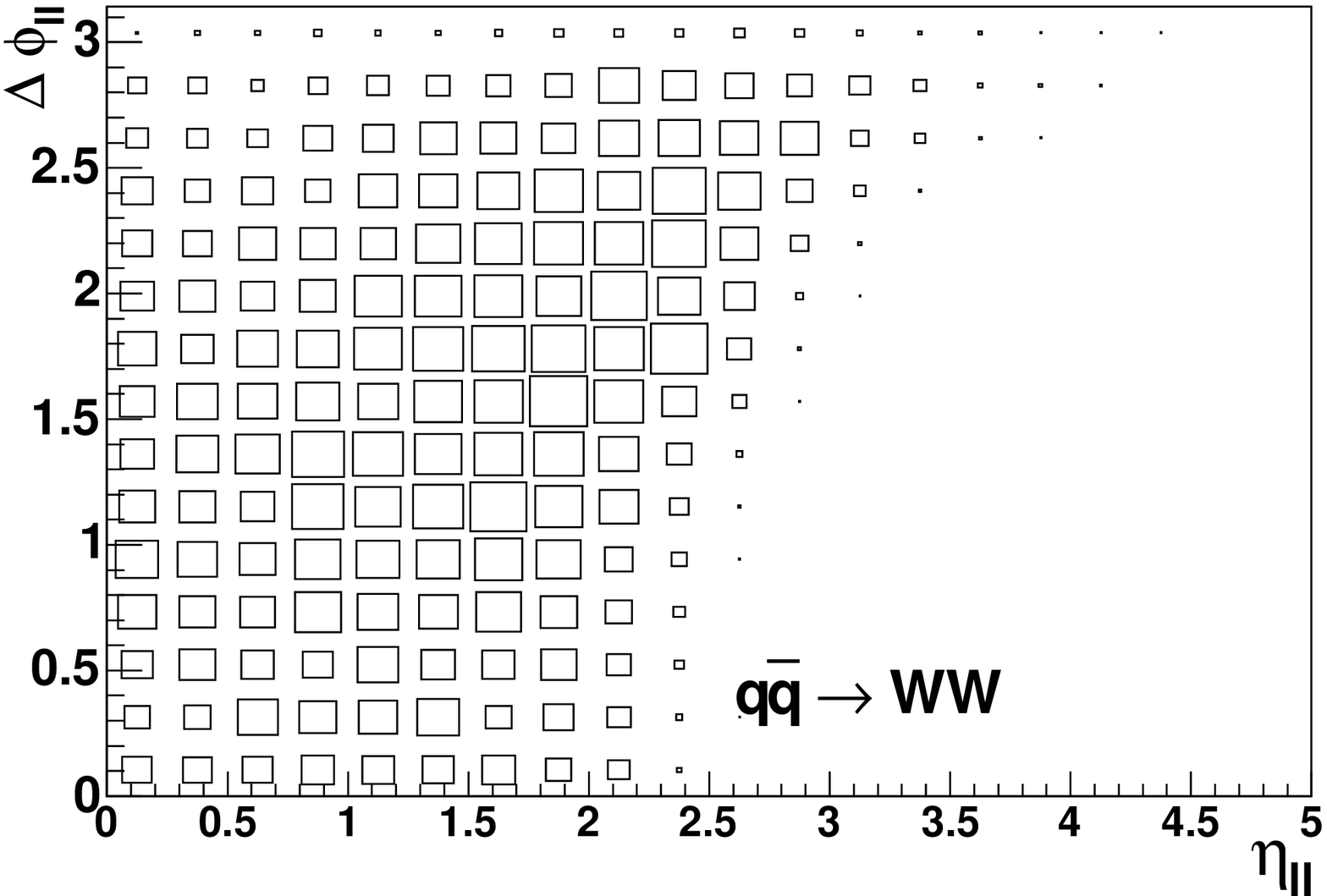,width=0.42\textwidth}}
\mbox{\epsfig{file=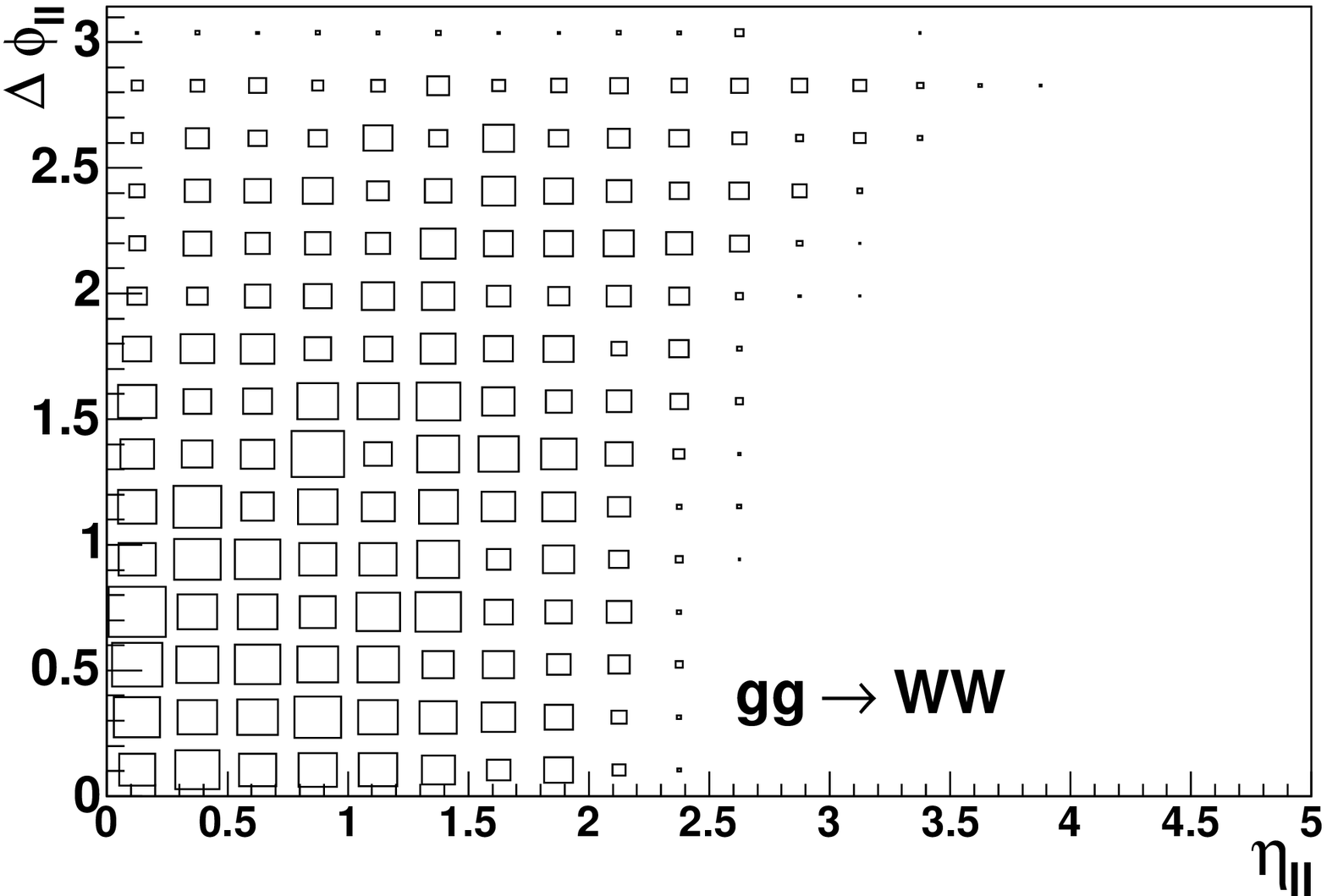,width=0.42\textwidth}}
\caption{\small \it
Distributions of the azimuthal separation \delphill\ versus \etall\ for signal events
with $m_H$ = 170~\Gcsit\ and for the \ttbar, \qqWW\ and \ggWW\ backgrounds.}
\label{f:delphi_eta}
\end{center}
\end{minipage}
\end{figure*}

Further cuts on the transverse mass $M_T$ of the dilepton-\ptmiss\ system, on the
pseudorapidity of the dilepton system \etall\ and on the azimuthal separation between
the two leptons
\delphill\ are applied to improve the signal-to-background ratio. The
motivation for these cuts is illustrated in Figs.~\ref{f:mt_delphi} and
\ref{f:delphi_eta}, where
the transverse mass and the pseudorapidity \etall\ are shown versus
\delphill\ for signal events with $m_H$ = 170~\Gcs\ and for the various backgrounds
considered. These plots show a striking difference between the \qqWW\ and the
\ggWW\ backgrounds. Whereas the \qqWW\ background peaks at large values of
\delphill\ and \etall, the \ggWW\ distributions are more signal
like, {\em i.e.}, a significant fraction of the \ggWW\ background has both
small \delphi\ and small \etall\ values.

\begin{table*}
\begin{center}
\footnotesize
\begin{tabular}{l||c ||c|c|c||c }
\hline
\hline
 & signal (fb) & \multicolumn{4}{c}{background (fb)} \\
 & $m_H = 170$ $\Gcsit$ & $qq \to WW$ & $gg \to WW$ & $\ttbar$ & Total \\
\hline
Lepton acceptance + \ptmiss\ & 230.9 & 825.9 & 113.8 & 11698 & 12638 \\
(cuts 1+2) & & & & & \\
+ Jet vetos, anti \ttbar\ and & 122.9 & 535.2 & 67.7 & 102.6 & 705.5 \\
anti $Z$ cuts (cuts 3-6) & & & & & \\
\hline
+ cut (7) & 113.4 & 363.7 & 51.9 & 72.0 & 487.6 \\
+ $\etall $ (cut 8) & 76.9 & 190.9 & 33.6 & 40.6 & 265.1 \\
+ Transverse mass (cut 9) & 21.5 & 24.4 &  7.0 & 8.7 & 40.1 \\
+ $\delphill$ (cut 10) & 19.3 & 11.6 & 4.1 & 2.6 & 18.3 \\
\hline
\hline
\end{tabular}
\vspace{0.5cm}
\caption{\small \it Accepted signal (for $m_H =$ 170~\Gcsit) and background cross-sections
 in fb for the $H \rightarrow WW \rightarrow \ell \nu \ell \nu$ (with $\ell = e, \mu)$ 
channel after the application of successive cuts. The contributions from $W \to \tau \nu \to \ell \nu \nu \nu$ 
are included.}
\label{t:acc}
\end{center}
\end{table*}

The additional cuts applied in the signal selection (cuts 7-10) aim to reduce
in particular the $WW$ background. After all cuts,
the signal-to-background ratio is found to be at the level of 1.
For an integrated luminosity of 10 \fbs\ at the LHC about 193 signal events and
a total background of 183 events are expected. The
background is still dominated by  $WW$ events, which account for a fraction
of about 86\%. About 26\% of the $WW$ background results from the
\ggWW\ process. The increase in the fraction of the \ggWW\ background
is due to the different behaviour of the \ggWW\ and \qqWW\ backgrounds
in the \delphill\ and \etall\ distributions, as discussed above. The final
signal selection cuts favour the \ggWW\ component.

\begin{figure*}
\begin{minipage}{1.0\linewidth}
\begin{center}
\mbox{\epsfig{file=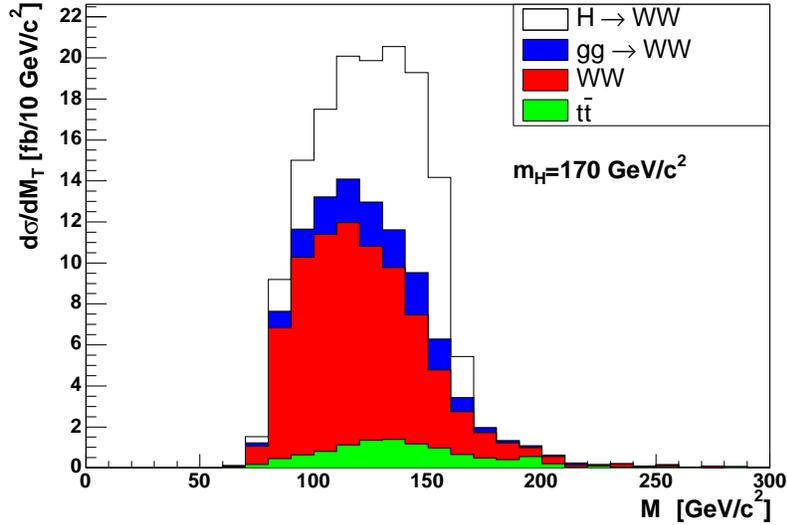,height=7.0cm}}
\caption{\small \it
Distribution of the transverse mass $M_T$ for a Higgs boson signal with $M_H$ = 170~\Gcsit\ and
backgrounds. The accepted cross section $d \sigma / d M_T$ (in fb / 10~\Gcsit) including
efficiency and acceptance factors is shown.}
\label{f:mt}
\end{center}
\end{minipage}
\end{figure*}

Due to the presence of neutrinos in the final state, it is not possible to reconstruct
a Higgs boson mass peak. Evidence for a Higgs boson signal has therefore to be extracted
from an excess of events above the backgrounds.
Unlike in the case of a narrow Higgs boson resonance, the background cannot
be determined in the experiment from sideband distributions.
The difficulty of a signal extraction is illustrated in Fig.~\ref{f:mt},
where the distribution of the reconstructed transverse mass is shown for a
Higgs boson signal with $m_H$ = 170~\Gcs\ and the backgrounds after
all cuts, except the transverse mass cut (cut 9).

\begin{figure*}
\begin{minipage}{1.0\linewidth}
\begin{center}
\mbox{\epsfig{file=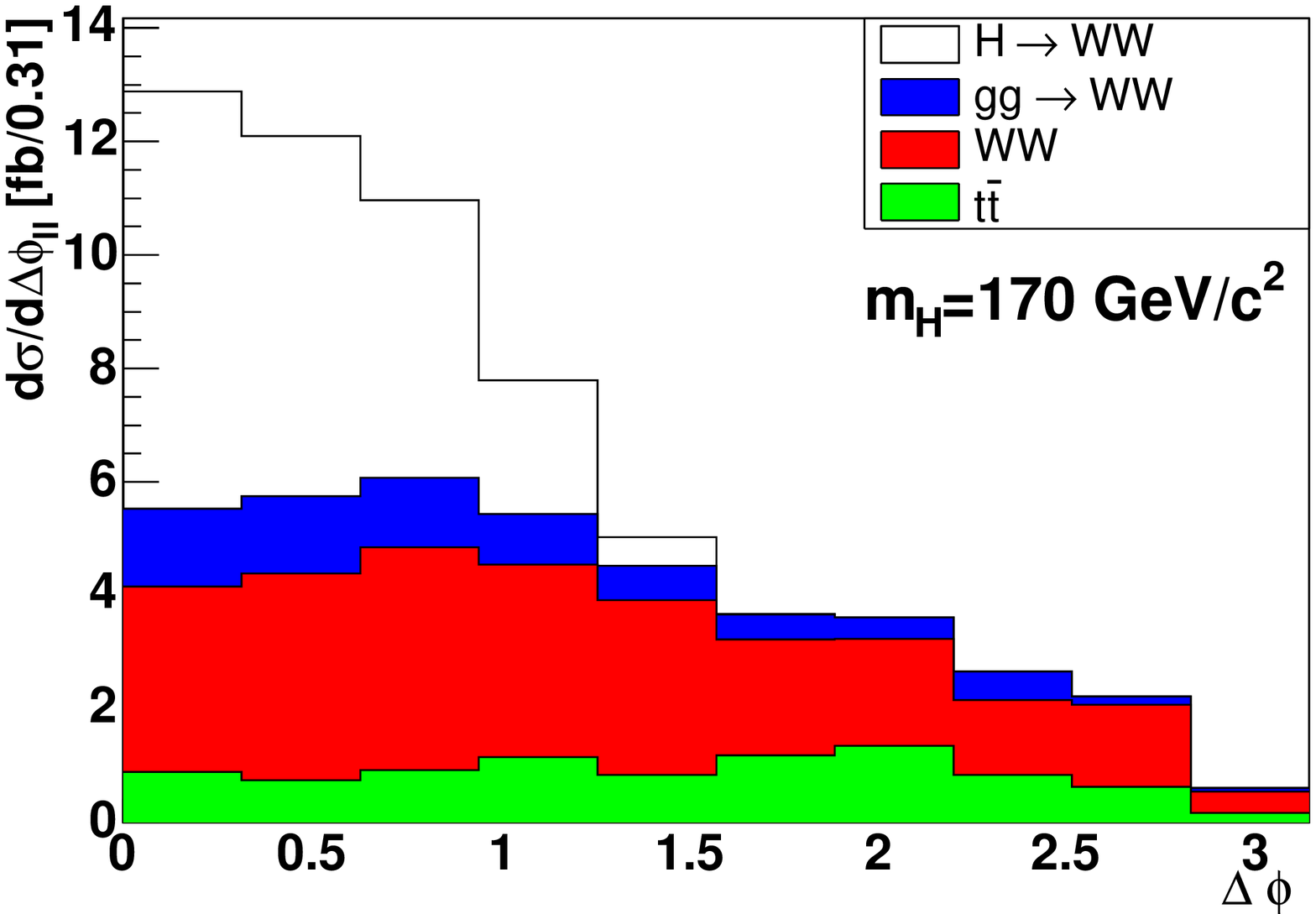,width=7.5cm}}
\mbox{\epsfig{file=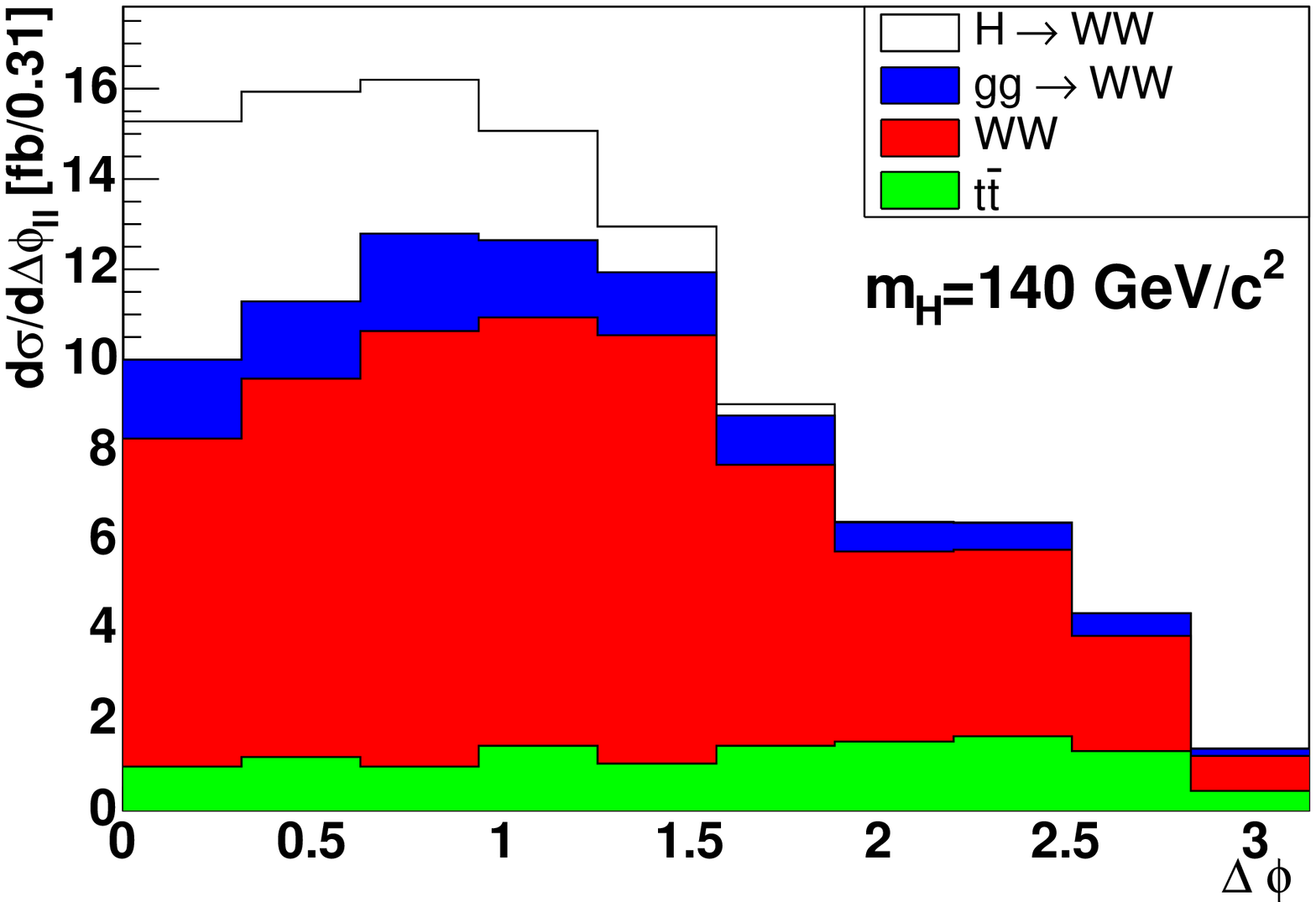,width=7.5cm}}
\caption{\small \it
Distribution of the azimuthal separation \delphill\ between the two leptons
after all cuts for backgrounds and Higgs bosons signals with $m_H$ = 170~\Gcs(left) and
$m_H$ = 140~\Gcs(right).}
\label{f:delphi}
\end{center}
\end{minipage}
\end{figure*}

However, it has been shown in several experimental studies
\cite{dreiner,atlas-vbf,han-turcot} that in particular the
distribution of the azimuthal separation \delphill\ can be used to constrain
and to normalize the background.
The dominant $qq \to WW$ and \ttbar\ backgrounds, which have been considered so far,
show after the application of selection cuts a rather flat behaviour.
Therefore, it was hoped that, after a normalization of the backgrounds in the
region of $\delphill > 1.5$, a reliable extrapolation of the background
could be performed with the help of Monte Carlo calculations.
Unfortunately, this situation is changed if the \ggWW\ background is included,
since this background has a large component which shows a signal-like behaviour.
Primarily it leads to an increased $WW$ background of the order of 35\%.
In addition, due to potentially large uncertainties on the shape of
the \ggWW\ background, the extrapolation uncertainties and therefore the
systematic uncertainties on the background estimates in the signal region increase.
The \delphill\ distribution found after the application of all cuts is shown in
Fig.~\ref{f:delphi} for two Higgs boson masses ($m_H$ = 170~\Gcs\ and $m_H$ = 140~\Gcs).

\section{Conclusions}

The production of $W$ pairs from the one-loop gluon fusion process is found to
significantly increase the $WW$ background in the Higgs search in
$H \to WW \to \ell \nu \ \ell \nu$ final states at the LHC. Since for this background the
correlations of the final state leptons are similar to those of the
Higgs boson signal, the fraction of the \ggWW\ contribution is enhanced by
the selection cuts foreseen in the experimental searches, {\em e.g.}, for
a Higgs boson with a mass of 170\,\Gcs, the $WW$ background is increased
by 35\% due to the \ggWW\ contribution, after final cuts are applied.
This larger background decreases the sensitivity of the LHC experiments for Higgs boson searches
in the $gg \to H \to WW$ mode. Due to the presence of two neutrinos in the final state
no mass peak can be reconstructed and it is necessary to normalize the background
in a phase space region where the signal contribution is small,
for example, using the distribution of the azimuthal difference between the leptons.
Due to the large \ggWW\ contribution in the signal region,
this normalization and subsequent extrapolation into
the signal region might be affected with larger uncertainties.
They lead in turn to larger systematic uncertainties on the background and thereby
reduce the signal significance further.

It should be stressed that the experimental collaborations should include the \ggWW\
background in their detailed detector simulation and re-evaluate the signal
significance in the $gg \to H \to WW$  mode. Given the striking differences
between the leading order \qqWW\ and the one-loop gluon-fusion process \ggWW\
in the final state lepton correlations, it appears necessary to use a full NLO
calculation in form of an event-by-event Monte Carlo generator for a more reliable
estimate of the Higgs boson discovery potential in this channel.

\acknowledgments

During the final phase of our work we became aware that also the group of Ref.~\cite{binoth}
was working on the same process. We compared our results with theirs and with the
older literature. The amplitudes are in agreement with each other. M.D. wants to thank 
David Rainwater for many useful discussions.


\clearpage

\appendix
\section{Helicity amplitudes}
In this appendix we give the helicity amplitudes needed for the
process $gg\to W^+W^-$\footnote{The full Monte Carlo program including
  the helicity amplitudes can be
  obtained by email from the authors {\em
    Michael.Duehrssen@physik.uni-freiburg.de,
  Peter.Marquard@durham.ac.uk}}.
We limit ourselves to present the results for
the case of one massless and one heavy quark with mass $m_1$. The
helicity configurations are defined for all incoming particles. In total
there are 8 independent amplitudes. The others can be obtained by bose
symmetry and parity transformation, e.g.
\begin{eqnarray*}
M_{++--}(\beta)&=&M_{++++}(\beta),\\
M_{+++-}(\beta)&=&M_{++-+}(\beta),\\
M_{+---}(\beta)&=&M_{+-++}(\beta),\\
M_{+--+}(\beta)&=&M_{+-+-}(-\beta),\\
M_{+++0}(\beta)=M_{++0+}(\beta)&=&M_{++-0}(-\beta)=M_{++0-}(-\beta),\\
M_{+-+0}(\beta)=M_{+-0+}(-\beta)&=&M_{+--0}(-\beta)=M_{+-0-}(\beta) .
\end{eqnarray*}
where
\begin{equation*}
  \beta = -\sqrt{1-4\frac{M_W^2}{s}} .
\end{equation*}
To obtain shorter expressions we introduce several new functions
\begin{align*}
s_1 &= s - M_W^2, &&
s_2 = s - 2 M_W^2, &&
s_4 = s - 4 M_W^2, \\
t_1 &= t - M_W^2, &&
t_2 = t - 2 M_W^2, \\
u_1 &= u - M_W^2, &&
u_2 = u - 2 M_W^2, \\
Y &= u t-M_W^4, &&
\Delta = i \sqrt{\frac{Y}{2 s_4 M_W^2}}.
\end{align*}
The helicity amplitudes are given in terms of the one loop scalar
integrals which are denoted as follows
\begin{align*}
  B_0(s)&=B(s,m_1^2,m_1^2) \\
  B_1(s)&=B(s,m_1^2,0) \\
  B_2(s)&=B(s,0,0)\\
  C_0(s)&=C(0,0,s,m_1^2,m_1^2,m_1^2) \\
  C_1^1(s) &= C(s,M_W^2,M_W^2,m_1^2,m_1^2,0) \\
  C_1^2(s) &= C(0,s,M_W^2,m_1^2,m_1^2,0) \\
  C_2^1(s) &= C(s,M_W^2,M_W^2,0,0,m_1^2) \\
  C_2^2(s) &= C(0,s,M_W^2,0,0,m_1^2) \\
  D_1(s,t) &= D(0,0,M_W^2,M_W^2,s,t,m_1^2,m_1^2,m_1^2,0) \\
  D_2(s,t) &= D(0,M_W^2,0,M_W^2,s,t,m_1^2,m_1^2,0,0) \\
  D_3(s,t) &= D(0,0,M_W^2,M_W^2,s,t,0,0,0,m_1^2)
\end{align*}
Here we followed the notation given in \cite{Hahn:1998yk}.
\paragraph{$M_{++++}$}
\begin{align*}
&M_{++++}=\frac{16\,( D_1(s,u) + D_1(s,t) + D_2(t,u) ) \,
      {m_1}^6}{s_4} \\& +
   \frac{2\,{m_1}^2 }{\beta\,s\,s_4\,t_1\,u_1}( 8\,C_1^2(u)\,{M_W}^2\,
         s\,( s + ( -2 + \beta ) \,s_2 ) \,
         t_1 + ( 8\,C_1^2(t)\,{M_W}^2\,s\,
            ( s + ( -2 + \beta ) \,s_2 ) \\&+
           t_1\,( \beta\,
               ( 16\,C_0(s)\,{M_W}^2\,s +
                 s_2\,
                  ( 4\,D_1(s,u)\,{M_W}^2\,s +
                    4\,D_1(s,t)\,{M_W}^2\,s \\&+
                    D_2(t,u)\,
                     ( s^2 - 2\,s\,s_2 -
                       {( t - u ) }^2 ) ) ) +
              ( s - 2\,s_2 ) \,
               ( 8\,C_0(s)\,s +
                 4\,D_1(s,u)\,{M_W}^2\,s \\&+
                 4\,D_1(s,t)\,{M_W}^2\,s +
                 D_2(t,u)\,s^2 - 2\,D_2(t,u)\,s\,s_2 -
                 D_2(t,u)\,t^2 + 2\,D_2(t,u)\,t\,u \\&-
                 D_2(t,u)\,u^2 ) ) ) \,u_1
        ) \\&-
   \frac{1} {2\,\beta\,
      s^2\,s_4\,{t_1}^2\,{u_1}^2}(4\,B_1(M_W^2)\,{M_W}^2\,s\,
       ( s + ( -2 + \beta ) \,s_2 ) \,
       ( s^4 - 2\,s^3\,s_2 +
         4\,s^2\,{( t - u ) }^2 \\&-
         2\,s\,s_2\,{( t - u ) }^2 -
         {( t - u ) }^4 ) +
      4\,B_1(u)\,s\,( -s + 2\,s_2 -
         \beta\,s_2 ) \,{t_1}^2\,
       ( s^3 - 2\,{( t - u ) }^2\,u \\&+
         2\,s^2\,( t + 2\,u ) +
         s\,( t^2 + 2\,t\,u + 5\,u^2 ) ) -
      16\,\beta\,{M_W}^2\,s^2\,t_1\,
       ( s^2 - 2\,s\,s_2 - {( t - u ) }^2 )
         \,u_1 \\&+ 8\,s^2\,( s - 2\,s_2 ) \,
       t_1\,( s^2 - 2\,s\,s_2 +
         {( t - u ) }^2 ) \,u_1 +
      4\,C_1^2(t)\,( s +
         ( -2 + \beta ) \,s_2 ) \,
       {t_1}^3\,( s^2 - 2\,s\,s_2 \\&-
         {( t - u ) }^2 ) \,{u_1}^2 +
      4\,C_2^2(t)\,( s +
         ( -2 + \beta ) \,s_2 ) \,
       {t_1}^3\,( s^2 - 2\,s\,s_2 -
         {( t - u ) }^2 ) \,{u_1}^2 \\&+
      D_2(t,u)\,( s +
         ( -2 + \beta ) \,s_2 ) \,
       {t_1}^2\,( s^2 - 2\,s\,s_2 -
         {( t - u ) }^2 ) \,
       ( s^2 - 2\,s\,s_2 + {( t - u ) }^2 )
         \,{u_1}^2 \\&+ 4\,B_1(t)\,s\,
       ( -s + 2\,s_2 - \beta\,s_2 ) \,
       ( s^3 - 2\,t\,{( t - u ) }^2 +
         2\,s^2\,( 2\,t + u ) +
         s\,( 5\,t^2 + 2\,t\,u + u^2 ) ) \,{u_1}^2
       \\&+ 4\,C_1^2(u)\,( s +
         ( -2 + \beta ) \,s_2 ) \,
       {t_1}^2\,( s^2 - 2\,s\,s_2 -
         {( t - u ) }^2 ) \,{u_1}^3 \\&+
      4\,C_2^2(u)\,( s +
         ( -2 + \beta ) \,s_2 ) \,
       {t_1}^2\,( s^2 - 2\,s\,s_2 -
         {( t - u ) }^2 ) \,{u_1}^3) \\&-
   \frac{4\,{m_1}^4 }{\beta\,s\,
      s_4}\,( 2\,
         ( D_1(s,u) + D_1(s,t) ) \,s\,
         ( s - 2\,s_2 ) +
        \beta\,( 8\,C_0(s)\,s + 2\,D_1(s,u)\,s^2 \\&+
           2\,D_1(s,t)\,s^2 + D_2(t,u)\,s^2 -
           2\,D_2(t,u)\,s\,s_2 + D_2(t,u)\,t^2 +
           4\,C_1^2(t)\,t_1 +
           4\,C_2^2(t)\,t_1 \\&- 2\,D_2(t,u)\,t\,u +
           D_2(t,u)\,u^2 + 4\,C_1^2(u)\,u_1 +
           4\,C_2^2(u)\,u_1 ) )
\end{align*}

\paragraph{$M_{+++-}$}
\begin{align*}
&M_{+++-}=- \frac{1}{s^2\,
        s_4\,{t_1}^2\,{u_1}^2}(( s^2 - 2\,s\,s_2 +
          {( t - u ) }^2 ) \,
        ( -2\,B_1(M_W^2)\,{M_W}^2\,s\,
           ( s^3 + s^2\,s_2 \\&- s\,{( t - u ) }^2 +
             s_2\,{( t - u ) }^2 ) +
          4\,B_1(u)\,{M_W}^2\,s\,{t_1}^2\,
           ( s - 2\,u ) +
          u_1\,( -2\,{M_W}^2\,s\,
              ( 4\,B_1(t)\,t \\&+
                D_2(t,u)\,s_2\,{t_1}^2 ) \,
              u_1 + s^2\,
              ( -4\,s_2\,t_1 +
                {M_W}^2\,
                 ( 4\,B_1(t) +
                   D_2(t,u)\,{t_1}^2 ) \,u_1
                ) \\&+ {M_W}^2\,{t_1}^2\,
              u_1\,( D_2(t,u)\,
                 {( t - u ) }^2 +
                4\,( C_1^2(t)\,t_1 +
                   C_2^2(t)\,t_1 +
                   ( C_1^2(u) + C_2^2(u) ) \,
                    u_1 ) ) ) ) ) \\&-
   \frac{4\,( D_1(s,u) + D_1(s,t) + D_2(t,u) ) \,
      {m_1}^6\,( s^2 - 2\,s\,s_2 -
        {( t - u ) }^2 ) }{s_4\,Y} \\&+
    \frac{{m_1}^4 }{s\,
      s_4\,Y}\,( 8\,C_0(s)\,s\,
         ( s^2 - 2\,s\,s_2 - {( t - u ) }^2 )
            + 4\,C_1^2(t)\,t_1\,
         ( s^2 - 2\,s\,s_2 - {( t - u ) }^2 )
            \\&+ 4\,C_2^2(t)\,t_1\,
         ( s^2 - 2\,s\,s_2 - {( t - u ) }^2 )
            \\&+ D_2(t,u)\,( 3\,s^2 - 6\,s\,s_2 -
           {( t - u ) }^2 ) \,
         ( s^2 - 2\,s\,s_2 + {( t - u ) }^2 )
            \\&+ 4\,D_1(s,u)\,s\,
         ( s^3 - 3\,s_2\,{( t - u ) }^2 +
           s^2\,( 5\,t + u ) +
           s\,( 7\,t^2 + 2\,t\,u - u^2 ) ) \\&+
        4\,D_1(s,t)\,s\,( s^3 -
           3\,s_2\,{( t - u ) }^2 +
           s^2\,( t + 5\,u ) +
           s\,( -t^2 + 2\,t\,u + 7\,u^2 ) ) \\&+
        4\,C_1^2(u)\,( s^2 - 2\,s\,s_2 -
           {( t - u ) }^2 ) \,u_1 +
        4\,C_2^2(u)\,( s^2 - 2\,s\,s_2 -
           {( t - u ) }^2 ) \,u_1 ) \\&+
   \frac{{m_1}^2\ }{2\,s\,s_4\,
      t_1\,u_1\,Y},( C_1^2(u)\,t_1\,
         ( 3\,s^5 + 2\,s_2\,{( t - u ) }^4 -
           s\,{( t - u ) }^3\,( 5\,t + 3\,u ) +
           2\,s^4\,( 5\,t + 7\,u ) \\&+
           2\,s^3\,( 5\,t^2 + 14\,t\,u - 3\,u^2 ) -
           4\,s^2\,u\,( -5\,t^2 + 10\,t\,u + 11\,u^2 ) ) +
        16\,C_1^1(s)\,s^2\,{( s - 2\,s_2 ) }^2\,
         t_1\,u_1 \\&-
        8\,C_0(s)\,s\,s_2\,t_1\,
         ( s^2 - 2\,s\,s_2 - {( t - u ) }^2 )
           \,u_1 - 4\,C_2^2(t)\,
         ( s - 2\,s_2 ) \,{t_1}^2\,
         ( s^2 - 2\,s\,s_2 + {( t - u ) }^2 )
           \,u_1 \\&- D_2(t,u)\,
         ( 2\,s - 3\,s_2 ) \,t_1\,
         {( s^2 - 2\,s\,s_2 + {( t - u ) }^2
             ) }^2\,u_1 \\&+
        C_1^2(t)\,( 3\,s^5 +
           2\,s_2\,{( t - u ) }^4 +
           s\,{( t - u ) }^3\,( 3\,t + 5\,u ) +
           2\,s^4\,( 7\,t + 5\,u ) \\&-
           4\,s^2\,t\,( 11\,t^2 + 10\,t\,u - 5\,u^2 ) +
           s^3\,( -6\,t^2 + 28\,t\,u + 10\,u^2 ) ) \,
         u_1 \\&+ 2\,D_1(s,u)\,s\,t_1\,
         ( s^4 - 4\,s^3\,s_2 +
           {( t^2 - u^2 ) }^2 +
           2\,s^2\,( 3\,t^2 + 4\,t\,u - u^2 ) \\&+
           4\,s\,( t^3 + t^2\,u - 3\,t\,u^2 - 3\,u^3 ) ) \,
         u_1 \\&+ 2\,D_1(s,t)\,s\,t_1\,
         ( s^4 - 4\,s^3\,s_2 -
           2\,s^2\,( t^2 - 4\,t\,u - 3\,u^2 ) +
           {( t^2 - u^2 ) }^2 \\&-
           4\,s\,( 3\,t^3 + 3\,t^2\,u - t\,u^2 - u^3 ) ) \,
         u_1 \\&- 4\,C_2^2(u)\,
         ( s - 2\,s_2 ) \,t_1\,
         ( s^2 - 2\,s\,s_2 + {( t - u ) }^2 )
           \,{u_1}^2 )
\end{align*}

\paragraph{$M_{+++0}$}
\begin{align*}
  &M_{+++0}=\frac{2\,( -1 + \beta ) \,\Delta\,{M_W}^2\, }{\beta\,s^2\,{t_1}^2\,
      {u_1}^2}
      ( -4\,B_1(M_W^2)\,{M_W}^2\,s\,
         ( -3\,s^2 + {( t - u ) }^2 ) \,
         ( t - u ) \\&+
        4\,B_1(u)\,s\,{t_1}^2\,
         ( s^2 + 2\,( t - u ) \,u +
           s\,( t + 3\,u ) ) -
        u_1\,( 4\,B_1(t)\,s^3\,u_1 -
           2\,s\,( 4\,B_1(t)\,t \\&+
              D_2(t,u)\,s_2\,{t_1}^2 ) \,
            ( t - u ) \,u_1 +
           {t_1}^2\,( t - u ) \,u_1\,
            ( D_2(t,u)\,{( t - u ) }^2 +
              4\,( C_1^2(t)\,t_1 +
                 C_2^2(t)\,t_1 \\&+
                 ( C_1^2(u) + C_2^2(u) ) \,
                  u_1 ) ) +
           s^2\,( u\,( 8\,t_1 +
                 4\,B_1(t)\,u_1 -
                 D_2(t,u)\,{t_1}^2\,u_1 ) \\&+
              t\,( -8\,t_1 +
                 12\,B_1(t)\,u_1 +
                 D_2(t,u)\,{t_1}^2\,u_1 )
              ) ) ) \\&- \frac{8\,
      ( D_1(s,u) + D_1(s,t) + D_2(t,u) ) \,
      \Delta\,{m_1}^6\,( t - u ) }{\beta\,
      Y} \\&+ \frac{2\,\Delta\,{m_1}^4\, }{\beta\,s\,
      Y}
      ( -( D_1(s,u)\,s\,
           ( 2\,s^2 + 2\,( -s_2 +
                i \,( -1 + \beta ) \,s_2 ) \,
              ( t - u ) \\&+
             s\,( ( 3 + i +
                   ( 1 - i ) \,\beta ) \,t +
                ( 5 - i - ( 1 - i ) \,\beta )
                   \,u ) ) ) \\&+
        D_1(s,t)\,s\,( 2\,s^2 -
           2\,( -s_2 +
              i \,( -1 + \beta ) \,s_2 ) \,
            ( t - u ) +
           s\,( ( 5 - i - ( 1 - i ) \,\beta
                 ) \,t \\&+ ( 3 + i +
                 ( 1 - i ) \,\beta ) \,u ) )
            + ( t - u ) \,
         ( 8\,C_0(s)\,s - i \,D_2(t,u)\,s^2 \\&+
           ( 1 + i ) \,\beta\,D_2(t,u)\,s\,
            ( s - 2\,s_2 ) +
           2\,i \,D_2(t,u)\,s\,s_2 +
           D_2(t,u)\,t^2 + 4\,C_1^2(t)\,t_1 +
           4\,C_2^2(t)\,t_1 \\&- 2\,D_2(t,u)\,t\,u +
           D_2(t,u)\,u^2 + 4\,C_1^2(u)\,u_1 +
           4\,C_2^2(u)\,u_1 ) ) \\&- \frac{( -1 + \beta ) \,\Delta\,
      {m_1}^2\,}{2\,
      \beta\,s\,t_1\,u_1\,Y}( 8\,C_0(s)\,s\,
         ( ( -1 + i ) \,s -
           2\,i \,s_2 ) \,t_1\,
         ( t - u ) \,u_1 \\&+
        ( 4 + 4\,i ) \,C_2^2(t)\,
         ( s - 2\,s_2 ) \,{t_1}^2\,
         ( t - u ) \,u_1 \\&+
        D_2(t,u)\,( ( 3 + i ) \,s -
           ( 4 + 2\,i ) \,s_2 ) \,
         t_1\,( s^2 - 2\,s\,s_2 +
           {( t - u ) }^2 ) \,( t - u ) \,
         u_1 \\&- 2\,D_1(s,u)\,s\,
         ( ( -1 + i ) \,s -
           2\,i \,s_2 ) \,t_1\,
         ( s^2 - 2\,s\,s_2 + t^2 - u^2 ) \,
         u_1 \\&+ 2\,D_1(s,t)\,s\,
         ( ( -1 + i ) \,s -
           2\,i \,s_2 ) \,t_1\,
         ( s^2 - 2\,s\,s_2 - t^2 + u^2 ) \,
         u_1 \\&- C_1^2(t)\,
         ( 8\,s^4 + s^3\,( 23\,t + 25\,u ) +
           2\,s_2\,( t - u ) \,
            ( t^2 + 4\,i \,{t_1}^2 - 2\,t\,u + u^2 )
            \\&+ 4\,s^2\,( 2\,t^2 + 7\,t\,u + 7\,u^2 ) -
           s\,( t - u ) \,
            ( 5\,t^2 + 4\,i \,{t_1}^2 + 14\,t\,u +
              13\,u^2 ) ) \,u_1 \\&+
        ( 4 + 4\,i ) \,C_2^2(u)\,
         ( s - 2\,s_2 ) \,t_1\,
         ( t - u ) \,{u_1}^2 \\&+
        C_1^2(u)\,t_1\,
         ( 8\,s^4 + s^3\,( 25\,t + 23\,u ) +
           4\,s^2\,( 7\,t^2 + 7\,t\,u + 2\,u^2 ) \\&-
           2\,s_2\,( t - u ) \,
            ( t^2 - 2\,t\,u + u^2 + 4\,i \,{u_1}^2 )
            + s\,( t - u ) \,
            ( 13\,t^2 + 14\,t\,u + 5\,u^2 +
              4\,i \,{u_1}^2 ) ) )
\end{align*}

\paragraph{$M_{++00}$}
\begin{align*}
&M_{++00}=\frac{8\,( D_1(s,u) + D_1(s,t) + D_2(t,u) ) \,
      {m_1}^6\,s}{{M_W}^2\,s_4} \\&-
   \frac{2\,{m_1}^4\ }{
      {M_W}^2\,s_4},( 8\,C_0(s)\,s +
        4\,D_1(s,t)\,{M_W}^2\,s +
        8\,D_1(s,t)\,{M_W}^2\,s_2 \\&+
        4\,D_1(s,u)\,{M_W}^2\,
         ( s + 2\,s_2 ) +
        4\,C_1^2(t)\,t_1 +
        4\,C_2^2(t)\,t_1 - 9\,D_2(t,u)\,t\,u +
        3\,D_2(t,u)\,t_2\,u \\&+
        4\,C_1^2(u)\,u_1 +
        4\,C_2^2(u)\,u_1 +
        3\,D_2(t,u)\,t\,u_2 -
        D_2(t,u)\,t_2\,u_2 ) \\&-
   \frac{8\,{M_W}^2\ }{s^2\,s_4\,{t_1}^2\,{u_1}^2},( B_1(M_W^2)\,{M_W}^2\,s\,
         ( s^4 - 2\,s^3\,s_2 +
           4\,s^2\,{( t - u ) }^2 -
           2\,s\,s_2\,{( t - u ) }^2 -
           {( t - u ) }^4 ) \\&-
        B_1(u)\,s\,{t_1}^2\,
         ( s^3 - 2\,{( t - u ) }^2\,u +
           2\,s^2\,( t + 2\,u ) +
           s\,( t^2 + 2\,t\,u + 5\,u^2 ) ) \\&-
        u_1\,( B_1(t)\,s\,
            ( s^3 - 2\,t\,{( t - u ) }^2 +
              2\,s^2\,( 2\,t + u ) +
              s\,( 5\,t^2 + 2\,t\,u + u^2 ) ) \,
            u_1 \\&- 4\,t_1\,Y\,
            ( -2\,s^2 + t_1\,u_1\,
               ( C_1^2(t)\,t_1 +
                 C_2^2(t)\,t_1 +
                 C_1^2(u)\,u_1 +
                 C_2^2(u)\,u_1 -
                 D_2(t,u)\,Y ) ) )
        )
    \\&+ \frac{2\,{m_1}^2\, }{{M_W}^2\,s\,
      s_4\,t_1\,u_1}
      ( C_1^2(u)\,{M_W}^2\,t_1\,
         ( 7\,s^3 + 2\,s_2\,{( t - u ) }^2 +
           4\,s^2\,( 3\,t + 4\,u ) \\&+
           s\,( 3\,t^2 + 18\,t\,u + 11\,u^2 ) ) \\&+
        u_1\,( C_1^2(t)\,{M_W}^2\,
            ( 7\,s^3 + 2\,s_2\,{( t - u ) }^2 +
              4\,s^2\,( 4\,t + 3\,u ) +
              s\,( 11\,t^2 + 18\,t\,u + 3\,u^2 ) ) \\&+
           4\,t_1\,( C_0(s)\,s\,
               ( s^2 - 2\,s\,s_2 + 2\,{s_2}^2
                 ) + {M_W}^2\,
               ( 2\,D_1(s,u)\,{M_W}^2\,s\,
                  s_2 +
                 2\,D_1(s,t)\,{M_W}^2\,s\,s_2 \\&+
                 C_2^2(t)\,s_4\,t_1 +
                 C_2^2(u)\,s_4\,u_1 +
                 3\,D_2(t,u)\,s\,Y -
                 4\,D_2(t,u)\,s_2\,Y )
              ) ) )
\end{align*}
            
\paragraph{$M_{+-+0}$}


\paragraph{$M_{+-+-}$}
%

\paragraph{$M_{+-00}$}
\begin{align*}
&M_{+-00}=\frac{4\,( D_1(s,u) + D_1(s,t) + D_3(s,t) +
        D_3(s,u) + 2\,D_2(t,u) ) \,{m_1}^8\,s^2}
      {{M_W}^2\,s_4\,Y} \\&-
   \frac{2\,{m_1}^6\,s\,}{{M_W}^2\,s_4\,Y}
      ( 4\,C_0(s)\,s - 4\,C_3(s)\,s +
        3\,D_1(s,u)\,s^2 + 3\,D_1(s,t)\,s^2 -
        2\,D_3(s,t)\,s^2 \\&- 2\,D_3(s,u)\,s^2 +
        D_2(t,u)\,s^2 - 2\,D_3(s,t)\,{s_2}^2 -
        2\,D_3(s,u)\,{s_2}^2 +
        ( 3 + i ) \,D_1(s,u)\,s\,t \\&-
        ( 1 - i ) \,D_1(s,t)\,s\,t +
        ( 1 + i ) \,D_3(s,t)\,s\,t -
        ( 3 - i ) \,D_3(s,u)\,s\,t +
        2\,i \,D_2(t,u)\,s\,t \\&- D_1(s,u)\,t^2 -
        D_1(s,t)\,t^2 - 3\,D_2(t,u)\,t^2 -
        ( 1 + i ) \,D_1(s,u)\,s\,u +
        ( 3 - i ) \,D_1(s,t)\,s\,u \\&-
        ( 3 + i ) \,D_3(s,t)\,s\,u +
        ( 1 - i ) \,D_3(s,u)\,s\,u -
        2\,i \,D_2(t,u)\,s\,u - 6\,D_1(s,u)\,t\,u \\&-
        6\,D_1(s,t)\,t\,u - 10\,D_2(t,u)\,t\,u -
        D_1(s,u)\,u^2 - D_1(s,t)\,u^2 - 3\,D_2(t,u)\,u^2
        ) \\&+
   \frac{{m_1}^4\, }{2\,
      {M_W}^2\,s_4\,Y}( -16\,C_1^2(t)\,s\,t_1\,
         t_2 - 16\,C_2^2(t)\,s\,t_1\,
         t_2 + 8\,C_0(s)\,s\,
         ( 8\,{M_W}^2\,s_1 +
           i \,s\,( t - u ) ) \\&-
        4\,C_1^1(s)\,s\,( s^2 - 2\,s\,s_2 -
           {( t - u ) }^2 ) -
        4\,C_2^1(s)\,s\,( s^2 - 2\,s\,s_2 -
           {( t - u ) }^2 ) \\&+
        D_3(s,u)\,s\,( ( 1 + i ) \,s^3 -
           4\,{s_2}^2\,
            ( ( 1 + i ) \,t +
              ( 5 - i ) \,u ) -
           2\,s^2\,( ( 1 + i ) \,t +
              ( 9 - 3\,i ) \,u ) \\&+
           s\,( ( -7 - 7\,i ) \,t^2 -
              ( 42 - 6\,i ) \,t\,u -
              ( 23 - i ) \,u^2 ) ) \\&+
        D_3(s,t)\,s\,( ( 1 - i ) \,s^3 -
           4\,{s_2}^2\,
            ( ( 5 + i ) \,t +
              ( 1 - i ) \,u ) -
           2\,s^2\,( ( 9 + 3\,i ) \,t +
              ( 1 - i ) \,u ) \\&-
           s\,( ( 23 + i ) \,t^2 +
              ( 42 + 6\,i ) \,t\,u +
              ( 7 - 7\,i ) \,u^2 ) ) \\&+
        D_1(s,t)\,( ( 7 + i ) \,s^4 +
           4\,s^3\,( 2\,i \,t + ( 4 - i ) \,u )
               - 4\,{( t^2 - u^2 ) }^2 +
           s^2\,( ( 7 + 5\,i ) \,t^2 \\&-
              ( 22 - 6\,i ) \,t\,u +
              ( 7 - 11\,i ) \,u^2 ) +
           2\,s\,( ( 5 + 3\,i ) \,t^3 -
              ( 5 + i ) \,t^2\,u -
              ( 13 - i ) \,t\,u^2 -
              ( 3 + 3\,i ) \,u^3 ) ) \\&+
        D_1(s,u)\,( ( 7 - i ) \,s^4 +
           4\,s^3\,( ( 4 + i ) \,t - 2\,i \,u )
               - 4\,{( t^2 - u^2 ) }^2 +
           s^2\,( ( 7 + 11\,i ) \,t^2 \\&-
              ( 22 + 6\,i ) \,t\,u +
              ( 7 - 5\,i ) \,u^2 ) +
           2\,s\,( ( -3 + 3\,i ) \,t^3 -
              ( 13 + i ) \,t^2\,u -
              ( 5 - i ) \,t\,u^2 +
              ( 5 - 3\,i ) \,u^3 ) ) \\&+
        8\,C_3(s)\,s\,( 2\,s^2 + 2\,{s_2}^2 +
           s\,( -2\,s_2 + i \,( -t + u )
              ) ) -
        16\,C_1^2(u)\,s\,u_1\,u_2 \\&-
        16\,C_2^2(u)\,s\,u_1\,u_2 +
        8\,D_2(t,u)\,( 4\,{s_2}^2 +
           s\,( -3\,s_2 + i \,( -t + u )
              ) ) \,Y ) \\&+
   \frac{2\,{M_W}^2\, }{
      s_4\,{t_1}^2\,{u_1}^2\,Y}( 2\,C_1^1(s)\,{t_1}^2\,
         ( s_2\,
            ( -s^2 + 2\,s\,s_2 +
              {( t - u ) }^2 ) +
           2\,i \,s\,s_4\,( t - u ) ) \,
         {u_1}^2 \\&- D_1(s,t)\,s\,t\,{t_1}^2\,
         ( s^2 - 2\,s\,s_2 + 5\,t^2 -
           6\,i \,t\,t_1 +
           2\,i \,t_1\,t_2 + 2\,t\,u + u^2 ) \,
         {u_1}^2 \\&+ 2\,C_1^2(t)\,{t_1}^3\,
         ( s^2 - 2\,s\,s_2 + 5\,t^2 -
           6\,i \,t\,t_1 +
           2\,i \,t_1\,t_2 + 2\,t\,u + u^2 ) \,
         {u_1}^2 \\&+ 2\,C_0(s)\,s\,{t_1}^2\,
         ( s^2 - 2\,s\,s_2 + 3\,t^2 +
           2\,i \,s_2\,( t - u ) + 2\,t\,u + 3\,u^2
           ) \,{u_1}^2 \\&+
        2\,C_3(s)\,s\,{t_1}^2\,
         ( s^2 - 2\,s\,s_2 + 3\,t^2 + 2\,t\,u + 3\,u^2 +
           2\,i \,s_2\,( -t + u ) ) \,
         {u_1}^2 \\&- D_3(s,t)\,s\,t\,{t_1}^2\,
         ( s^2 - 2\,s\,s_2 + 5\,t^2 -
           2\,i \,t_1\,t_2 + u^2 +
           2\,t\,( 3\,i \,t_1 + u ) ) \,
         {u_1}^2 \\&+ 2\,C_2^2(t)\,{t_1}^3\,
         ( s^2 - 2\,s\,s_2 + 5\,t^2 -
           2\,i \,t_1\,t_2 + u^2 +
           2\,t\,( 3\,i \,t_1 + u ) ) \,
         {u_1}^2 \\&+ 2\,C_2^1(s)\,{t_1}^2\,
         ( -( s^2\,s_2 ) +
           s_2\,{( t - u ) }^2 +
           2\,s\,( {s_2}^2 +
              i \,s_4\,( -t + u ) ) )
          \,{u_1}^2 \\&-
        D_1(s,u)\,s\,{t_1}^2\,u\,{u_1}^2\,
         ( s^2 - 2\,s\,s_2 + t^2 + 2\,t\,u + 5\,u^2 +
           6\,i \,u\,u_1 -
           2\,i \,u_1\,u_2 ) \\&+
        2\,C_1^2(u)\,{t_1}^2\,{u_1}^3\,
         ( s^2 - 2\,s\,s_2 + t^2 + 2\,t\,u + 5\,u^2 +
           6\,i \,u\,u_1 -
           2\,i \,u_1\,u_2 ) \\&-
        D_3(s,u)\,s\,{t_1}^2\,u\,{u_1}^2\,
         ( s^2 - 2\,s\,s_2 + t^2 + 2\,t\,u + 5\,u^2 -
           6\,i \,u\,u_1 +
           2\,i \,u_1\,u_2 ) \\&+
        2\,C_2^2(u)\,{t_1}^2\,{u_1}^3\,
         ( s^2 - 2\,s\,s_2 + t^2 + 2\,t\,u + 5\,u^2 -
           6\,i \,u\,u_1 +
           2\,i \,u_1\,u_2 ) \\&-
        4\,B_1(u)\,{t_1}^2\,
         ( s^2 - 2\,s\,s_2 + t^2 + 2\,t\,u + 5\,u^2 ) \,
         Y \\&+ 2\,B_1(M_W^2)\,
         ( s^4 - 2\,s^3\,s_2 -
           6\,s\,s_2\,{( t - u ) }^2 +
           4\,s^2\,( t^2 + u^2 ) +
           {( t - u ) }^2\,( 3\,t^2 + 2\,t\,u + 3\,u^2 )
           ) \,Y \\&-
        4\,B_1(t)\,( s^2 - 2\,s\,s_2 + 5\,t^2 +
           2\,t\,u + u^2 ) \,{u_1}^2\,Y -
        32\,t_1\,u_1\,{Y}^2 )\\
&+ \frac{{m_1}^2\, }{4\,
      {M_W}^2\,s\,s_4\,t_1\,u_1\,
      Y}( 4\,C_1^1(s)\,s\,t_1\,
         ( ( 2\,s^2 - s\,s_2 -
              2\,{s_2}^2 ) \,
            ( s^2 - 2\,s\,s_2 - {( t - u ) }^2
              ) \\& 2\,i \,s^2\,s_4\,( t - u )
           ) \,u_1 \\&+
        4\,C_2^1(s)\,s\,t_1\,
         ( -( ( 2\,s^2 - 3\,s\,s_2 +
                2\,{s_2}^2 ) \,
              ( s^2 - 2\,s\,s_2 -
                {( t - u ) }^2 ) ) -
           2\,i \,s^2\,s_4\,( t - u ) ) \,
         u_1 \\&+ 4\,C_0(s)\,s^2\,t_1\,
         ( -3\,s^3 + 2\,s^2\,s_2 -
           2\,i \,( 2\,s^2 - 3\,s\,s_2 +
              2\,{s_2}^2 ) \,( t - u ) \\&+
           4\,s_2\,{( t - u ) }^2 -
           s\,( t^2 - 10\,t\,u + u^2 ) ) \,u_1 \\&+
        4\,C_3(s)\,s^2\,t_1\,
         ( s^3 + 6\,s^2\,s_2 + 8\,{s_2}^3 -
           2\,i \,( 2\,s^2 - 3\,s\,s_2 +
              2\,{s_2}^2 ) \,( t - u ) \\&-
           s\,( 13\,t^2 + 30\,t\,u + 13\,u^2 ) ) \,
         u_1 \\&+ 2\,D_1(s,t)\,s\,t_1\,
         ( ( -1 - i ) \,s^5 +
           s^4\,( ( 1 - 5\,i ) \,t -
              ( 2 + 2\,i ) \,u ) -
           2\,{s_2}^2\,
            ( s_2 \\&+ i \,( t - u ) ) \,
            {( t - u ) }^2 -
           2\,s^3\,( ( 2 + 2\,i ) \,t^2 -
              ( 5 - 7\,i ) \,t\,u -
              ( 1 + i ) \,u^2 ) -
           s^2\,( ( 1 + 3\,i ) \,t^3 \\&+
              ( 2 + 6\,i ) \,t^2\,u -
              ( 19 - 15\,i ) \,t\,u^2 -
              ( 8 + 8\,i ) \,u^3 ) +
           s\,s_2\,( ( -11 + i ) \,t^3 +
              ( 7 - 9\,i ) \,t^2\,u \\&-
              ( 5 - 15\,i ) \,t\,u^2 -
              ( 7 + 7\,i ) \,u^3 ) ) \,
         u_1 \\&+ 4\,C_2^2(t)\,{t_1}^2\,
         ( ( 1 - i ) \,s^4 -
           2\,s^3\,( ( 4 + 3\,i ) \,t - i \,u )
               + 2\,{( t^2 - u^2 ) }^2 -
           s^2\,( ( 13 + 5\,i ) \,t^2 \\&+
              ( 18 + 2\,i ) \,t\,u +
              ( 1 - 7\,i ) \,u^2 ) -
           2\,s\,( 3\,t^3 +
              2\,i \,{s_2}^2\,( t - u ) +
              9\,t^2\,u + 5\,t\,u^2 - u^3 ) ) \,u_1 \\&+
        2\,D_1(s,u)\,s\,t_1\,
         ( ( -1 + i ) \,s^5 -
           2\,{s_2}^2\,
            ( s_2 + i \,( t - u ) ) \,
            {( t - u ) }^2 +
           s^4\,( ( -2 + 2\,i ) \,t \\&+
              ( 1 + 5\,i ) \,u ) +
           2\,s^3\,( ( 1 - i ) \,t^2 +
              ( 5 + 7\,i ) \,t\,u -
              ( 2 - 2\,i ) \,u^2 ) +
           s\,s_2\,( ( -7 + 7\,i ) \,t^3 \\&-
              ( 5 + 15\,i ) \,t^2\,u +
              ( 7 + 9\,i ) \,t\,u^2 -
              ( 11 + i ) \,u^3 ) +
           s^2\,( ( 8 - 8\,i ) \,t^3 +
              ( 19 + 15\,i ) \,t^2\,u \\&-
              ( 2 - 6\,i ) \,t\,u^2 -
              ( 1 - 3\,i ) \,u^3 ) ) \,
         u_1 \\&+ 2\,D_3(s,t)\,s^2\,t_1\,
         ( ( 1 - i ) \,s^4 +
           ( 1 - i ) \,s^3\,( 3\,t + 4\,u ) +
           {s_2}^2\,( ( 17 + 7\,i ) \,t^2 \\&+
              ( 6 - 6\,i ) \,t\,u +
              ( 1 - i ) \,u^2 ) +
           2\,s^2\,( ( 10 + 2\,i ) \,t^2 +
              ( 7 - 7\,i ) \,t\,u +
              ( 3 - 3\,i ) \,u^2 ) \\&+
           s\,( ( 31 + 9\,i ) \,t^3 +
              ( 50 - 2\,i ) \,t^2\,u +
              ( 19 - 19\,i ) \,t\,u^2 +
              ( 4 - 4\,i ) \,u^3 ) ) \,
         u_1 \\&+ 2\,D_3(s,u)\,s^2\,t_1\,
         ( ( 1 + i ) \,s^4 +
           ( 1 + i ) \,s^3\,( 4\,t + 3\,u ) +
           2\,s^2\,( ( 3 + 3\,i ) \,t^2 +
              ( 7 + 7\,i ) \,t\,u \\&+
              ( 10 - 2\,i ) \,u^2 ) +
           {s_2}^2\,( ( 1 + i ) \,t^2 +
              ( 6 + 6\,i ) \,t\,u +
              ( 17 - 7\,i ) \,u^2 ) +
           s\,( ( 4 + 4\,i ) \,t^3 \\&+
              ( 19 + 19\,i ) \,t^2\,u +
              ( 50 + 2\,i ) \,t\,u^2 +
              ( 31 - 9\,i ) \,u^3 ) ) \,
         u_1 \\&+ C_1^2(t)\,
         ( s^6 + 2\,{s_2}^2\,{( t - u ) }^4 +
           10\,s^5\,( 3\,t + u ) +
           4\,s^4\,( 13\,t^2 + 27\,t\,u + 8\,u^2 ) \\&+
           2\,s\,{( t - u ) }^2\,
            ( 7\,t^3 + 17\,t^2\,u + 17\,t\,u^2 + 7\,u^3 ) +
           4\,s^3\,( 19\,t^3 + 22\,t^2\,u + 35\,t\,u^2 \\&+ 12\,u^3 )
               + s^2\,( 81\,t^4 + 48\,t^3\,u + 18\,t^2\,u^2 +
              72\,t\,u^3 + 37\,u^4 ) -
           4\,i \,s\,{t_1}^2\,
            ( s^3 + s^2\,( 6\,t \\&- 2\,u ) +
              4\,{s_2}^2\,( t - u ) +
              s\,( 5\,t^2 + 2\,t\,u - 7\,u^2 ) ) ) \,
         u_1 \\&+ 4\,C_2^2(u)\,t_1\,
         ( ( 1 + i ) \,s^4 -
           2\,s^3\,( i \,( t - 3\,u ) + 4\,u ) +
           2\,{( t^2 - u^2 ) }^2 -
           s^2\,( ( 1 + 7\,i ) \,t^2 \\&+
              ( 18 - 2\,i ) \,t\,u +
              ( 13 - 5\,i ) \,u^2 ) -
           2\,s\,( -t^3 + 2\,i \,{s_2}^2\,
               ( t - u ) + 5\,t^2\,u + 9\,t\,u^2 + 3\,u^3 )
           ) \,{u_1}^2 \\&+
        C_1^2(u)\,t_1\,
         ( s^6 + 2\,{s_2}^2\,{( t - u ) }^4 +
           10\,s^5\,( t + 3\,u ) +
           4\,s^4\,( 8\,t^2 + 27\,t\,u + 13\,u^2 ) \\&+
           2\,s\,{( t - u ) }^2\,
            ( 7\,t^3 + 17\,t^2\,u + 17\,t\,u^2 + 7\,u^3 ) +
           4\,s^3\,( 12\,t^3 + 35\,t^2\,u + 22\,t\,u^2 \\&+ 19\,u^3 )
               + s^2\,( 37\,t^4 + 72\,t^3\,u + 18\,t^2\,u^2 +
              48\,t\,u^3 + 81\,u^4 ) +
           4\,i \,s\,( s^3 - 2\,s^2\,( t - 3\,u ) \\&+
              4\,{s_2}^2\,( -t + u ) +
              s\,( -7\,t^2 + 2\,t\,u + 5\,u^2 ) ) \,
            {u_1}^2 ) \\&-
        64\,D_2(t,u)\,{M_W}^2\,s_2\,
         t_1\,u_1\,{Y}^2 )
       \end{align*}

\end{document}